\documentclass[twocolumn,prl]{revtex4}

\usepackage{graphicx}

\newcommand{\nn}{\nonumber}

\begin{document}
\title{Direct CP Violation in $B\to \phi K_{s}$ and New Physics}

\author{Marco Ciuchini}
\affiliation{%
I.N.F.N. Sezione di Roma III and Dip. di Fisica, Universit\`a di Roma Tre\\
Via della Vasca Navale 84, I-00146 Rome, Italy
}%

\author{Luca Silvestrini}
\affiliation{
I.N.F.N. Sezione di Roma and Dip. di Fisica, Universit\`a di Roma 
``La Sapienza''\\
P.le A. Moro 2, I-00185 Rome, Italy
}%

\begin{abstract}
In the presence of large New Physics contributions to loop-induced
$b\to s$ transitions, sizable direct CP violation in $B\to \phi K$
decays is expected on general grounds. We compute explicitly
CP-violating effects using QCD factorization and find that, even in
the restricted case in which New Physics has the same penguin
structure as the Standard Model, the rate asymmetry can be of order
one. We briefly discuss a more general scenario and comment on the
inclusion of power-suppressed corrections to factorization.
\end{abstract}


\maketitle

With the advent of $B$ factories, the measurement of CP asymmetries in
non-leptonic $B$ decays has emerged as a very powerful probe of New
Physics (NP) beyond the Standard Model (SM). It was pointed out a few
years ago that the comparison of time-dependent CP asymmetries in
different decay channels measuring the same weak phase in the SM could
provide evidence of NP in $B$ decay
amplitudes~\cite{Grossman:1996ke}. In particular, $a_{CP}(B\to J/\Psi
K_{s})$ and $a_{CP}(B\to \phi K_{s})$ both measure $\sin2\beta$ with
negligible hadronic uncertainties in the
SM~\cite{Grossman:1997gr}. However, $B\to \phi K_{s}$, being a pure
penguin process, is expected to be much more sensitive to NP than the
tree-level dominated $B\to J/\Psi K_{s}$ decay.  In many explicit
examples of NP, it has been shown that sizable differences in these
two asymmetries can be
generated~\cite{Ciuchini:1997zp}--\cite{Datta:2002nr}. The first
measurements of $a_{CP}(B\to \phi K_{s})$ by the BaBar and Belle
collaborations display a $2.7 \sigma$ deviation from the observed
value of $a_{CP}(B\to J/\Psi K_{s})$~\cite{Aubert:2002nx,Abe:2002bx},
leaving open the possibility of a NP effect in $B\to \phi K_{s}$
\cite{nir:ichep}.

In this letter, we focus on the possibility of having direct CP
violation in $B\to \phi K_{s}$ in the presence of generic NP
contributions to the $b\to s\bar s s$ transition at the loop level.
For simplicity, we first illustrate our argument using QCD
factorization in the limit $m_b\to\infty$~\cite{Beneke:1999br},
neglecting electroweak corrections. Then, we briefly discuss possible
effects of power-suppressed terms.

We write the decay amplitude as
\begin{eqnarray}
\mathcal{A}(B\to \phi K_s)&=&-\frac{G_F}{\sqrt{2}} F_0^{B\to K} f_\phi
\sum_{i=3}^5 \left[ \lambda_u \tilde a^u_i +\lambda_c \tilde
a^c_i+\right.\nn\\ 
&&\left.\lambda_t \left(\tilde a^t_i+\tilde
a^\mathrm{NP}_i\right)\right]\,,
\label{eq:amp}
\end{eqnarray} 
where $\lambda_q=V_{qb}^*V_{qs}$, $F_0^{B\to K}$ is the semileptonic
$B\to K$ form factor evaluated at the $\phi$ mass and $f_\phi$ is the
$\phi$ decay constant. The coefficients $\tilde a^q_i$ are defined in
terms of the usual $a^q_{i,\mathrm{I}}$ and $a^q_{i,\mathrm{II}}$, 
introduced in QCD
factorization~\cite{Yang:2000xn,Beneke:2001ev}, as follows:
\begin{eqnarray}
&& \tilde a^u_{(3,5)}=0\,,~\tilde a^c_{(3,5)}=0\,,~\tilde
a^t_{(3,5)}=a_{(3,5),\mathrm{I}}+a_{(3,5),\mathrm{II}}\,,\nn\\ 
&& \tilde a^{(u,c)}_{4}=a^{(u,c)}_{4,\mathrm{I}}\,(C_{3,\dots,6}\to
0)\,,\nn\\ 
&&\tilde a^t_{4}=a^u_{4,\mathrm{I}}\,(C_{1,2}\to
0)+a_{4,\mathrm{II}}\,.
\end{eqnarray} 
The notation $a^u_{4,\mathrm{I}}\,(C_{1,2}\to 0)$ means that one has
to take the expression for $a^u_{4,\mathrm{I}}$ given in
ref.~\cite{Beneke:2001ev} neglecting terms proportional to $C_1$ and
$C_2$.  Furthermore, the coefficients $\tilde a^\mathrm{NP}_i$ in
eq.~(\ref{eq:amp}) account for the NP contributions and are defined as
$\tilde a^\mathrm{NP}_i=\tilde a^t_i\,(C_{3,\dots,6}\to
C^\mathrm{NP}_{3,\dots,6}/\lambda_t)$. For discussing NP
effects, it is useful  to distinguish the different $\lambda_q$ contributions,
without using the unitarity of the CKM matrix. In fact, terms
proportional to $\lambda_u$ and $\lambda_c$ are not modified by NP
loop effects. Since $\lambda_u$ is doubly Cabibbo suppressed with
respect to $\lambda_{c,t}$, we neglect it in the following discussion.

In Table~\ref{tab:atilde}, we report the values of the coefficients
$\tilde a^q_i$. It is remarkable that $\tilde a^c_4$ has comparable
real and imaginary parts and, correspondingly, a large strong phase
even in the infinite mass limit. However, $a^c_4=\tilde
a^c_4-\tilde a^t_4$, which enters the SM decay amplitude, has a smaller
strong phase, due to the constructive (destructive) interference in
the real (imaginary) parts. In other words, the strong phase of the SM
amplitude is accidentally smaller than its natural value
within QCD factorization. Notice, in addition, that $\vert\tilde
a^c_4\vert$ and $\vert\tilde a^t_4\vert$ are comparable in size.

\begin{table}
\caption{\label{tab:atilde} Numerical values of the coefficients
$\tilde a^q_i$ relevant to our discussion obtained for $\mu=m_b=4.2$
GeV, $\alpha_s(M_Z)=0.119$, $m_c(m_b)=1.3$ GeV. }
\begin{ruledtabular}
\begin{tabular}{lrr}
& Re & Im \\
\hline
$\tilde a^c_3$ & $0$  & $0$\\
$\tilde a^c_4$ & $-1.4\times 10^{-2}$  & $-1.1\times 10^{-2}$\\
$\tilde a^c_5$ & $0$ & $0$ \\
$\tilde a^t_3$ & $-4.3\times 10^{-3}$  & $-2.7\times 10^{-3}$ \\
$\tilde a^t_4$ &  $1.9\times 10^{-2}$ & $-3.4\times 10^{-3}$ \\
$\tilde a^t_5$ &  $4.1\times 10^{-3}$ & $3.1\times 10^{-3}$ \\
\end{tabular}
\end{ruledtabular}
\end{table}

Assuming that NP effects affect $C_{3,\dots,6}$, we can consider two
different scenarios: \textit{i)} a universal penguin-like contribution
parametrized as $C^\mathrm{NP}_{3,\dots,6}=\lambda_t r^\mathrm{NP}
e^{i\phi^\mathrm{NP}} C_{3,\dots,6}$; \textit{ii)} a general NP
contribution affecting $C_{3,\dots,6}$ in a non universal way. This
is the case, for example, of general $R_P$-conserving SUSY models
where, in addition to penguins, there is also a box
contribution~\cite{Gabbiani:1996hi}.

\begin{figure}
\includegraphics[width=8cm]{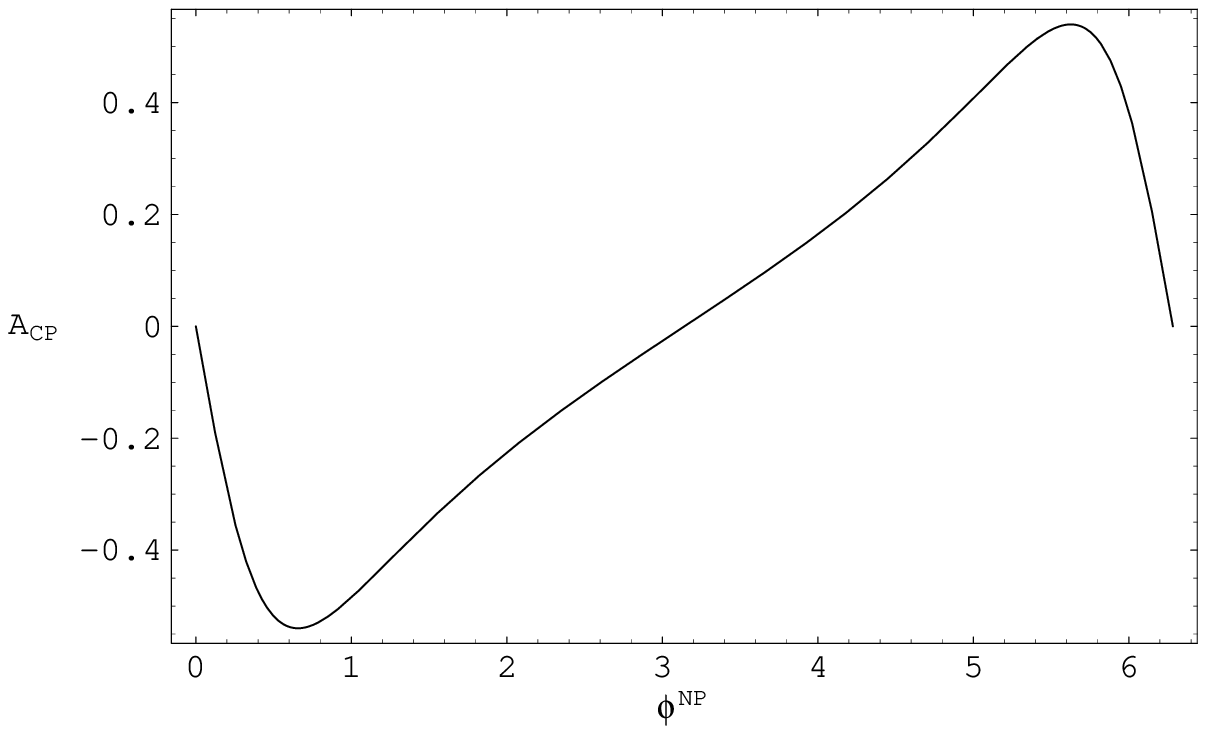}% Here is how to import EPS art
\caption{\label{fig:acp} The direct CP asymmetry in $B^0\to \phi K_s$ 
from eq.~\ref{eq:acp1}, for $r^\mathrm{NP}=1$, as a function of
$\phi^\mathrm{NP}$.}
\end{figure}

It is easy to see that, in both scenarios, there is more than one
contribution to the amplitude carrying different strong and weak
phases. Since the strong phases are not negligible, 
one expects sizable direct CP violation if the NP
contribution is large enough.  Indeed, in the first scenario, we have
\begin{eqnarray}
\mathcal{A}(B\to \phi K_s)&\simeq &-\frac{G_F}{\sqrt{2}} F_0^{B\to K}
f_\phi \sum_{i=3}^5 \left[ \lambda_c \tilde a^c_i+\right.\nn\\
&&\left.\lambda_t \left(1+r^\mathrm{NP}\,e^{i
\phi^\mathrm{NP}}\right)\tilde a^t_i\right]\,.
\end{eqnarray}  
Using the values in Table~\ref{tab:atilde}, we get
\begin{eqnarray}
\mathcal{A}(B\to \phi K_s)&\propto &\lambda_c (1.4+1.1\, i)+\nn\\
&&\lambda_t \left(1+r^\mathrm{NP}\,e^{i \phi^\mathrm{NP}}\right)
(-1.9+0.3\, i)\,.
\label{eq:acp1}
\end{eqnarray}  
It is apparent that, for $r^\mathrm{NP}$ of $\mathcal{O}(1)$, a large
rate asymmetry is generated, namely $\vert\bar
\mathcal{A}/\mathcal{A}\vert\equiv \vert \mathcal{A}(\bar B^0\to \phi K_s)\vert
/ \vert \mathcal{A}(B^0\to \phi K_s)\vert\neq 1$ (see fig.~\ref{fig:acp}). 
Correspondingly, the full
expression for the time-dependent asymmetry, including the $\cos\Delta
M_B t$ term, should be used and hadronic uncertainties are expected in
the extraction of the weak phases from the data.

In the more general scenario \textit{ii)}, there are even more terms
in the amplitude with different strong and weak phases. In this case,
$a^\mathrm{NP}_i$ contain an admixture of strong and weak
phases. Therefore, it is no longer useful to use the notation of
eq.~(\ref{eq:amp}).  As an example, we give the coefficients $C_{\phi
K}$ and $S_{\phi K}$ of the time-dependent CP asymmetry computed in a
SUSY model with $\mathcal{O}(1)$ $\tilde s-\tilde b$ mixing, for an
average squark and gluino mass of 250 GeV (see
refs.~\cite{Silvestrini:ichep,Ciuchini:fut} for a detailed
analysis). For central values of the parameters in QCD factorization
and the extreme value $(\delta^d_{23})_{LL}=e^{3 \pi i/2}$ 
(for the definition, see ref.~\cite{Gabbiani:1996hi}), we get
\begin{equation}
C_{\phi K}=-0.24\,,\qquad S_{\phi K}=-0.13\,.
\end{equation} 

To conclude our discussion, we notice that, as suggested by $B\to
K\pi$ decays, large corrections to QCD factorization in the infinite
$b$-mass limit are expected in penguin-dominated $b\to s$
decays~\cite{Ciuchini:2001gv}. However, the inclusion of power
corrections following any of the available
approaches~\cite{Beneke:2001ev},~\cite{Ciuchini:2001gv}--\cite{Ciuchini:2002} 
can only
strengthen our conclusion, since in general subleading terms produce
additional strong phases (barring accidental cancellations). 
Furthermore, given the dependence of
hadronic matrix elements on the final state, no simple relation among
the time-dependent CP asymmetries in $B\to \phi K_{s}$, $B\to
\eta^\prime K_{s}$, and other penguin-dominated $b\to s$ transitions
can be established. Therefore, it is quite possible that, in the
presence of NP, $a_{CP}(B\to \phi
K_{s}) \neq a_{CP}(B\to \eta^\prime K_{s})$, contrary to what very recently
suggested in ref.~\cite{nir:ichep}.
If the present $2.7\sigma$ discrepancy between
$S_{J/\Psi K}$ and $S_{\phi K}$ will be confirmed, pointing to a large
NP contribution in the $B\to \phi K_{s}$ decay amplitude, a
non-vanishing $C_{\phi K}$ is expected on general grounds, as well as
CP violation in the decay $B^+\to \phi K^+$.

We thank F. Zwirner for carefully reading the manuscript.

\end{document}